\documentclass{article}

\usepackage{arxiv}

\usepackage[utf8]{inputenc} 
\usepackage[T1]{fontenc}    
\usepackage{hyperref}       
\usepackage{url}            
\usepackage{booktabs}       
\usepackage{amsfonts}       
\usepackage{nicefrac}       
\usepackage{microtype}      
\usepackage{lipsum}		
\usepackage{subfig}
\usepackage{graphicx}
\usepackage{float}
\usepackage{amsmath}
\usepackage{booktabs}
\usepackage{multirow}
\usepackage{doi}

\DeclareMathOperator{\Tr}{Tr}

\title{PE-GAN: Prior Embedding GAN for PXD Images at Belle II}

\author{ \href{https://orcid.org/0000-0003-4095-9657}{\includegraphics[scale=0.06]{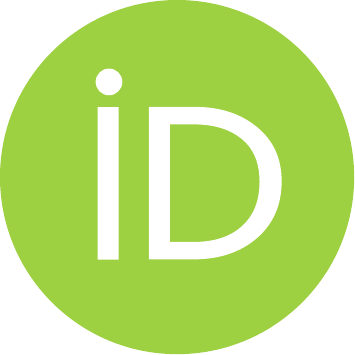}\hspace{1mm}Hosein Hashemi}\\
	Faculty of Physics\\
	Ludwig Maximilians University of Munich,\\
	Munich, Germany \\
	\texttt{gh.hashemi@physik.uni-muenchen.de} \\
	 \AND
	 Nikolai Hartmann \\
	 Faculty of Physics\\
	 Ludwig Maximilians University of Munich,\\
	 Munich, Germany \\
	 \texttt{nikolai.hartmann@physik.uni-muenchen.de} \\
	 \AND
	 Thomas Kuhr \\
	 Faculty of Physics\\
	 Ludwig Maximilians University of Munich,\\
	 Munich, Germany \\
	 \texttt{thomas.Kuhr@lmu.de} \\
  	 \AND
	 Martin Ritter \\
	 Faculty of Physics\\
	 Ludwig Maximilians University of Munich,\\
	 Munich, Germany \\
	 \texttt{martin.ritter@lmu.de} \\
     \AND
  	 Matej Srebre \\
	 Faculty of Physics\\
	 Ludwig Maximilians University of Munich,\\
	 Munich, Germany \\
	 \texttt{matej.srebre@physik.uni-muenchen.de} \\
}

\date{}

\renewcommand{\headeright}{CHEP 2021}
\renewcommand{\undertitle}{}

\hypersetup{
pdftitle={PE-GAN: Prior Embedding GAN for PXD images at Belle II},
pdfsubject={hep-ph, cs.LG, hep-ex, Data Analysis, Statistics and Probability},
pdfauthor={Hosein.~Hashemi},
pdfkeywords={GANs, High-granularity detector simulation, PXD Belle~II, Contrastive Learning},
}

\begin{document}
\maketitle

\begin{abstract}
	 The pixel vertex detector (PXD) is an essential part of the Belle II detector recording particle positions. Data from the PXD and other sensors allow us to reconstruct particle tracks and decay vertices. The effect of background hits on track reconstruction is simulated by adding measured or simulated background hit patterns to the hits produced by simulated signal particles. This model requires a large set of statistically independent PXD background noise samples to avoid a systematic bias of reconstructed tracks. However, data from the fine-grained PXD requires a substantial amount of storage. As an efficient way of producing background noise, we explore the idea of an on-demand PXD background generator using conditional Generative Adversarial Networks (GANs), adapted by the number of PXD sensors in order to both increase the image fidelity and produce sensor-dependent PXD hitmaps.
\end{abstract}


\section{Introduction}
\label{intro}
Belle II~\cite{b2} is a B-factory experiment located at the SuperKEKB~\cite{kekb} accelerator in Tsukuba, Japan operating at peak luminosity of $2\times10^{34}$ cm\textsuperscript{-2}s\textsuperscript{-1}. The accelerator collides electrons and positrons at a center of mass energy of $\sqrt{s}\approx 10.6$ GeV, which prompts to $\Upsilon(4S)$ resonance state that decays almost exclusively to entangled pairs of either charged or neutral $B$ mesons. These processes are called signal decays which are isotropic events. We also have other processes which are jet-like continuum events $(e^+e^- \longrightarrow q\Bar{q},~q = u, d, s, c)$. The precise measurement of $CP$-violation and the search for rare or “forbidden” decays of the B-meson and the tau-lepton as signals for New Physics relies heavily on a large number of recorded events and the precision with which B-meson and lepton decay vertices can be reconstructed. In order to do so, Belle II plans to collect a 50 times larger dataset than its predecessor, Belle. The accelerator's upgrade resulted in the desired increase of the collision rate of particles. Thus, we expect larger background at design luminosity for all sub-detectors. To profit from the much larger recorded dataset, the systematic uncertainties of simulated datasets must be well under control. This requires an accurate incorporation of beam background effects in the detector simulation. 
\par
Various background processes lead to signals in the detector that do not originate from the physics processes of interest. These processes can be categorised into two types, beam-induced and luminosity dependent processes. The beam-induced processes originate from the synchroton radiation and collisions of beam particles with residual gas in the beampipe, bending magnets or particles within a bunch, whereas luminosity dependent processes comprise electron-positron collisions leading to “non-interesting” physics processes such as Bhabha scattering or two-photon processes. 
\par
There are two methods to include and simulate these beam background effects, the background mixing method and the background overlay method. In the background mixing method the pre-simulated Geant4~\cite{geant} background hits are added to simulated energy deposit as SimHits from a signal event. Then, both background and signal contributions are merged and digitized at the same time. This has the advantage that the subsequent simulation of the digitization correctly takes into account cases where energy deposits from background and signal particles contribute to the same channel. On the other hand the comparison of background simulations using this so-called mixing technique with measurements show sizable discrepancies. The background overlay method uses random trigger events along with real data to create a background event sample. This works on digit level and thus needs a detector-speciﬁc treatment of cases where background and signal contribute to the same channel. Nevertheless it is expected to yield a more realistic simulation of beam background eﬀects. 

\section{Pixel Vertex Detector}
\label{sec-1}
The Pixel Vertex Detector (PXD)~\cite{pxd} was designed to be the innermost semi-conductor sub-detector of Belle II. The PXD measures the position of traversing particles originating from a particle collision in order to do precise reconstruction of decay vertices. As a result it is placed very closely to the interaction point, hence exposed to a huge amount of background radiation. The designed configuration of PXD consists of 40 modules within 2 layers of detector. The inner layer has 16 modules assembled in 8 planar ladders. The outer layer is composed from 12 ladders or 24 modules. Each module consists of a $250\times 768$ pixel matrix of DEPFET~\cite{depfet} pixel sensors which culminates in a $250\times 768$ hitmap. Currently only 20 modules are installed. With a readout time of $20~\mu s$ the PXD integrates over about $5000$ bunch crossings. Thus, the majority of pixels with a charge above threshold stems from the beam background. Pixel signals are only read out if their value exceeds a threshold. In our simulated data the threshold is at 7 ADU. We will address this in section \ref{sec-7}. 

\subsection{The Problem}
\label{sec-2}
A problem with the background overlay method is that large amounts of resources are required for the readout, storage, and distribution of the random trigger events. A single event in the format to be used for the background overlay has a size of about $200$ kB. This is roughly a factor 20 more than the size of events in the format used for analyses. Furthermore, the beam background files have to be distributed to all sites where run-dependent Monte-Carlo samples are produced. This multiplies the storage demands by the number of sites. In particular, smaller sites do not have the storage resources with the required size and bandwidth to host full-beam background datasets. Even if one copies the background files from the main server, and wants to access them through the network, each job has to access multiple files, for example for producing $1$ Tb of MC files, one has to copy $20$ Tb PXD background files which dramatically increases the data transfer and networking traffic.
\par
Our solution to this problem is to make it an end-to-end procedure. First, to generate these hitmaps on the fly and upon request for each sensor with a conditional generative agent, then extract PXD hits data from the fired pixels. The only storage cost would be the weights of the generative model.

\section{PXD Background Generation}
\label{sec-3}
This work is in the direction of the "Fast Simulation" campaign in Particle Physics that had been started after the introduction of the Generative Adversarial Networks (GAN), by Goodfellow et al.~\cite{gan} for image data. GAN is a flexible unsupervised deep learning architecture that can be trained to learn a function to match a given data distribution. GANs have been widely used in Particle Physics in order to have a faster detector simulation for LHC~\cite{lhc_1}~\cite{lhc_2}, but it has not been used for Pixel detector. GANs involve two networks, the Generator, and the Discriminator, in a two-player zero-sum game whose purpose, respectively, is to map random noise to samples and discriminate real and generated samples. In the end, the goal is to find a Nash equilibrium to a two-player min-max problem. In our case, we need a model to generate sensor-dependent PXD background hitmaps (images) as fast as possible with the highest fidelity. For the moment with the current approach, we ignore correlations between the two layers, since most of the background is coming from out-of-time events. As we usually do not have the corresponding SVD hits for two correlated PXD background hits caused by the same particles, so no track would be reconstructed. Thus, the expected effect on tracking performance from neglecting the correlations is small but still has to be studied in detail.
\par
Our data in each event consists of 40 sensor-dependent grey-scale $250\times 768$ images. For training, we used 40000 MC simulated events. And for evaluation, we have 10000 MC simulated events. The events are simulated for two times the expected background level at design luminosity. We are also planning to use the real data (random trigger) in the future as well.
\par
Here we are facing several challenges:
\begin{enumerate}
    \item Having high-resolution images. In general with GANs, there is no guarantee of balance between the training of the generator and discriminator. As a consequence, one network may be more powerful than the other, which in most cases is the discriminator. When the discriminator becomes much stronger than the generator, the fake images become too easy to be separated from real ones, thus reaching a point where gradients from the discriminator vanish, providing no guidance for further training of the generator. This happens more frequently when one is dealing with high-resolution images due to the diﬃculty of generating meaningful high-frequency details.  On the other hand, having less powerful discriminator results in a mode collapse, as with the generator greedily optimizing its loss function in order to “fool” the discriminator, each generator update is a partial collapse towards mode collapse~\cite{Metz}. A crucial aspect in maintaining a balance between solutions for vanishing Gradients and mode collapse is a discriminator with high generalization capabilities. Hence, in the end, it is provided by deeper networks with non-local blocks along with having PXD ladder numbers (sensor IDs) as conditional classes in an attempt to capture global and local information and dependencies which are notoriously memory consuming.
    
    \item Having sparse images. Convolutional Neural Network (CNN) as a part of the discriminator has trouble with sparse data~\cite{sparse}. They require dense data to learn well. Naively applying CNNs to sparse data only works partially as they are sensitive to missing data. Thus, we need more sophisticated methods and prior information such as sensor IDs to be injected into the learning procedure. 
    
    \item The lack of evaluation while training. When one deals with the generic image dataset such as ImageNet~\cite{imagenet}, one can use the Inception score (IS)~\cite{mbdis} or the Frechet Inception Score (FID)~\cite{fid} for training evaluation and validation to choose the best model as the GAN training is utterly unstable. But in our case, due to the specificity of our dataset, the commonly used evaluation metrics for natural images don't apply.
    
\end{enumerate}
\subsection{The Model}
\label{sec-4}
After considering a mixture of the models and structures, and tuning many hyperparameters by trial and error minimization based on our evaluation methods mentioned in \autoref{sec-5}, we chose a modified version of BigGAN-deep~\cite{biggan} with contrastive conditioning~\cite{contragan} and consistency regularization~\cite{crgan}. 

We started with adding class labels which were 40 sensor numbers to the model in an extra embedding layer to our former model, WGAN-gp~\cite{chep}, which acted as a baseline to our study. The problem was slow event-based image generation in comparison to the overall simulation time due to its large kernel filter size, and not capturing entirely the pixel intensity distribution. Then, we went through more complex but faster models due to the lower kernel size and special residual block which were designed for high-resolution image generation such as BigGAN~\cite{biggan}. During the training, our models undergo training collapse, thus early stopping was crucial. As suggested in BigGAN's paper~\cite{biggan}, we monitored and shown in Fig. \ref{fig-3}, the first singular values of each weight matrix for both the generator and the discriminator grow throughout training, but whenever we had mode collapse or discriminator overfitting to the training set, the first singular values explodes drastically for the generator. We tried many treatments, such as lowering the learning rate for the discriminator, using different regularization methods such as Orthogonal Regularization, Minibatch Discrimination~\cite{mbdis}, Dropout, and L2 regularization, Latent Optimization~\cite{logan}, Feature Quantisation~\cite{fqgan} did not work either. As a result, we thought that we have to go for a simpler model.
\par
Then, based on the context of our image dataset, and memory limits, along with the feedbacks and failures from the model during the training, we decided to use a simpler model. Thus, we used SAGAN~\cite{sagan}. Self-Attention GAN is an architecture that allows both the generator and discriminator to model long-range dependencies over the image manifold. The key idea is to make the generator able to "see" not only the convolutional kernel range but also the global details. Normally, Convolutions as local blocks have very nice properties such as parameter sharing and translation invariance and they learn in a hierarchical order. Nevertheless, when it comes to complex structures and elaborated geometrical contours, it requires long-range details. This is where attention enters. With this choice of architecture, we learned a lot about how the difference between the models would affect our result and the final evaluation result was surpassed by a deeper model with fewer intricacies with respect to BigGAN but more layers than SAGAN.
\par
In our model, we increased the depth of each block in both the generator and the discriminator, with 32 multiplication channels in each layer. After some training and ablation study over BigGAN-deep, we decided to remove the Projection Discrimination which was originally introduced in SNGAN~\cite{sngan}. Projection Discrimination's objective is to maximize the projection of the embeddings of real images on the corresponding target embeddings while minimizing the inner-product values when the images are fake. However, since it only utilizes data-to-class relation, it results in the discriminator’s overfitting and training collapse problems~\cite{biggan}. Hence, we replaced it with the contrastive conditioning discrimination mechanism~\cite{contragan}. It is based on a conditional contrastive loss that comes from metric learning that considers data-to-data relations in each batch. It pulls the multiple-image embeddings closer to each other in the embedding space when the class labels are the same. In other words, the discriminator can capture not only data-to-class but also data-to-data relations between samples. 

Within the context of metric learning loss, due to the fact that we are dealing with a close-conditional image generation task, we also added an embedding of the prior information along with the class labels to both the discriminator and generator. The relational embedding module in the discriminator consists of a linear projection of semantic information along with a self-attention over the concatenated version of label embeddings and semantic information projection. Let $\mathbf{X}=\{x_1,...,x_m\}$, where $x_i \in \mathbb{R}^{W\times H}$ be sampled training images and $\mathbf{y}=\{y_1,...,y_m\}$, where $y_i \in \mathbf{Z}$ be the corresponding labels from 1 to 40, and $\mathbf{s}=\{s_1,...,s_m\}$, where $s_i \in \mathbf{R}$ be the mean occupancies for each class of sensors. Also, we define a linear projection function $l^{(d)}:\mathbb{R}^k \rightarrow \mathbb{S}^d$ and an embedding function $\mathbf{e}:y_i \rightarrow \mathbb{R}^{d/4}$. For the relational embedding, we have 40 number of tokens of dimension $d$ as query, key, and value from the output of the concatenation, and a dot-product self-attention~\cite{sa} which calculates the weighted average of feature representations with the weight proportional to the dot-product similarity score between pairs of representation,
\[Q=l^{(d)}~[l^{(d/4)}(\mathbf{s})~||~\mathbf{e}]\]
\[A=softmax~(\frac{QK^T}{\sqrt{d}})~V.\]
Then, using the linear projection of the output of the discriminator, one can define the conditional contrastive loss are as, 
\[L(x_i,y_i,s_i;t)=-\log\left(\frac{\exp{(h_i^Ta_i/t)}+\sum^m_{k=1}\mathbf{1}_{k=i}.\exp{(h_i^Th_k/t)}}{\exp{(h_i^Ta_i/t)}+\sum^m_{k=1}\mathbf{1}_{k\neq i}.\exp{(h_i^Th_k/t)}}\right)
\] 
\[H=l^{(d)}[D(x_i)]\]

Where $a_i \in A$ and $h_i \in H$ . Minimizing the above contrastive loss will reduce distances between the embeddings of images with the same labels while maximizing the others. It considers the data-to-data relations $h_i^Th_k$ and data-to-class relations $h_i^Ta_i$ without augmentations. Thus, we help the discriminator to be guided by not only the image but also by the fine-grained prior information as an inductive bias injection. This prior semantic information is based on different mean occupancies from the simulated images. This extra information injection boosted both the performance regarding the mean-occupancy distribution and the stability of the training for almost 20000 more iterations. 

We also used consistency regularization~\cite{crgan} within our training which prolonged the training stability significantly, and delayed collapse. The idea behind consistency regularization is to penalize the sensitivity of the classiﬁer to some semantics-preserving noise. In our case, the noise perturbation is image ﬂipping and random shift as images with these augmented artifacts exist in the generator's outputs.

Eventually, our model, as depicted in Table~\ref{tab:1}, using the above technologies along with orthogonal initialized~\cite{ortho} residual blocks~\cite{biggan} for both the generator and the discriminator and Hinge loss~\cite{sagan}, succeeded in the sensor-dependent PXD hitmap generation. We trained our model with the PyTorch framework~\cite{pytorch} over Tesla V100 GPU. This model generates 40 hitmaps for an event in around 6 seconds on one CPU.

\begin{table}[ht]
\centering
\subfloat[Generator]{
\begin{tabular}{c} 
 \toprule
 \toprule
  $z\in \mathbb{R}^{128} \sim \mathcal{N}(0,I)$\\ Embed(y)~$\in\mathbb{R}^{128}$\\$ch=32$\\
 \hline
  Linear(128+128)~$\rightarrow4\times12\times16ch$\\
 \hline
 ResBlock~$16ch\rightarrow16ch$\\
 \hline
 ResBlock up~$16ch\rightarrow16ch$\\
 \hline
 ResBlock~$16ch\rightarrow16ch$\\
 \hline
 ResBlock up~$16ch\rightarrow8ch$\\
 \hline
 ResBlock~$8ch\rightarrow8ch$\\
 \hline
 ResBlock up~$8ch\rightarrow8ch$\\
 \hline
 ResBlock~$8ch\rightarrow8ch$\\
 \hline
 ResBlock up~$8ch\rightarrow4ch$\\
 \hline
 Non-Local Attention Block\\
 \hline
 ResBlock~$4ch\rightarrow4ch$\\
 \hline
 ResBlock up~$4ch\rightarrow2ch$\\
 \hline
 ResBlock~$2ch\rightarrow2ch$\\
 \hline
 ResBlock up~$2ch\rightarrow ch$\\
 \hline
 BN, ReLU, $3\times3$~Conv~$ch\rightarrow 1$\\
 \hline
 Tanh\\
 \bottomrule
 \bottomrule
\end{tabular}}\hspace{8mm}
\subfloat[Discriminator]{
\begin{tabular}{c} 
 \toprule
  \toprule
  zero padded gray-scale image\\$x\in\mathbb{R}^{256\times768\times1}$\\
 \hline
 ConvBlock~$1\rightarrow ch$\\
 \hline
 ResBlock down~$ch\rightarrow 2ch$\\
 \hline
 ResBlock~$2ch\rightarrow 2ch$\\
 \hline
 ResBlock down~$2ch\rightarrow4ch$\\
 \hline
 ResBlock~$4ch\rightarrow 4ch$\\
 \hline
 ResBlock down~$4ch\rightarrow8ch$\\
 \hline
 ResBlock~$8ch\rightarrow 8ch$\\
 \hline
 ResBlock down~$8ch\rightarrow8ch$\\
 \hline
 ResBlock~$8ch\rightarrow 8ch$\\
 \hline
 ResBlock down~$8ch\rightarrow16ch$\\
  \hline
 ResBlock~$16ch\rightarrow 16ch$\\
 \hline
 ResBlock down~$16ch\rightarrow16ch$\\
  \hline
 ResBlock~$16ch\rightarrow 16ch$\\
 \hline
 ReLU, Global sum pooling\\
 \hline
 Self-Attention~[Proj(Embed(y)~||~Linear(s))],\\ Norm(Linear512$\rightarrow1024$)\\
 \hline
 Conditional Contrastive loss\\
  \bottomrule
 \bottomrule
\end{tabular}}
\caption{\label{tab:1}The Architecture of the Generator and the Discriminator. "$ch$" denotes the number of multiplicative channels, "ResBlock*", the residual blocks introduced in BigGAN-deep, "BN", the batch normalization, "ReLU", the rectified linear unit, "Norm", a normalization operator, "Proj" and "Linear", a linear projection or an MLP.}
\end{table}

\section{Model Validation}
\label{sec-5}
For model evaluation we have two categories of methods. The ones that we use within the training process, and the ones after training for validation.
\begin{enumerate}
    \item During training:
    \begin{itemize}
        \item Image visualization (Turing Test)
        \item Monitoring singular values for Weight matrices
    \end{itemize}
    \item After training:
    \begin{itemize}
        \item Image pixel intensity analysis
        \item Sensor number occupancy distribution 
        \item Physics analysis over impact parameters resolutions 
    \end{itemize}
    
\end{enumerate}

\subsection{Training Evaluation Methods}
\label{sec-6}
\begin{figure}[ht]
  \centering
  \subfloat[a][real]{\includegraphics[width=0.5\textwidth,clip]{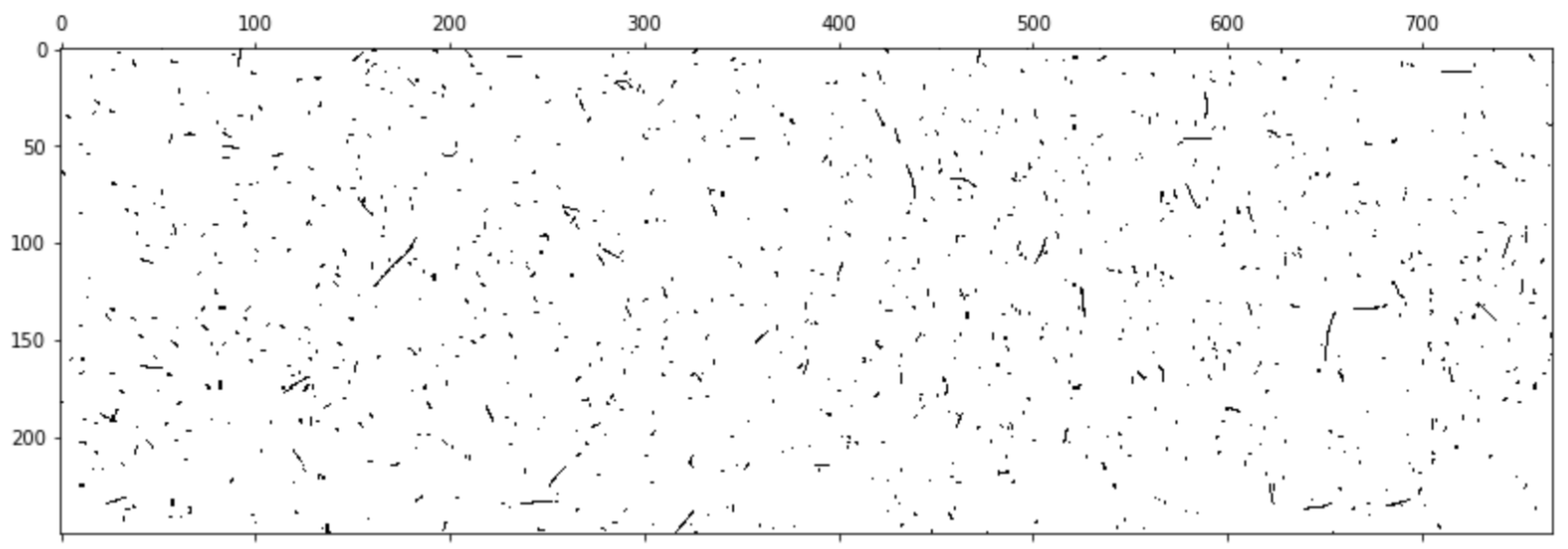} \label{fig:a}} \\
  \subfloat[b][generated]{\includegraphics[width=0.5\textwidth,clip]{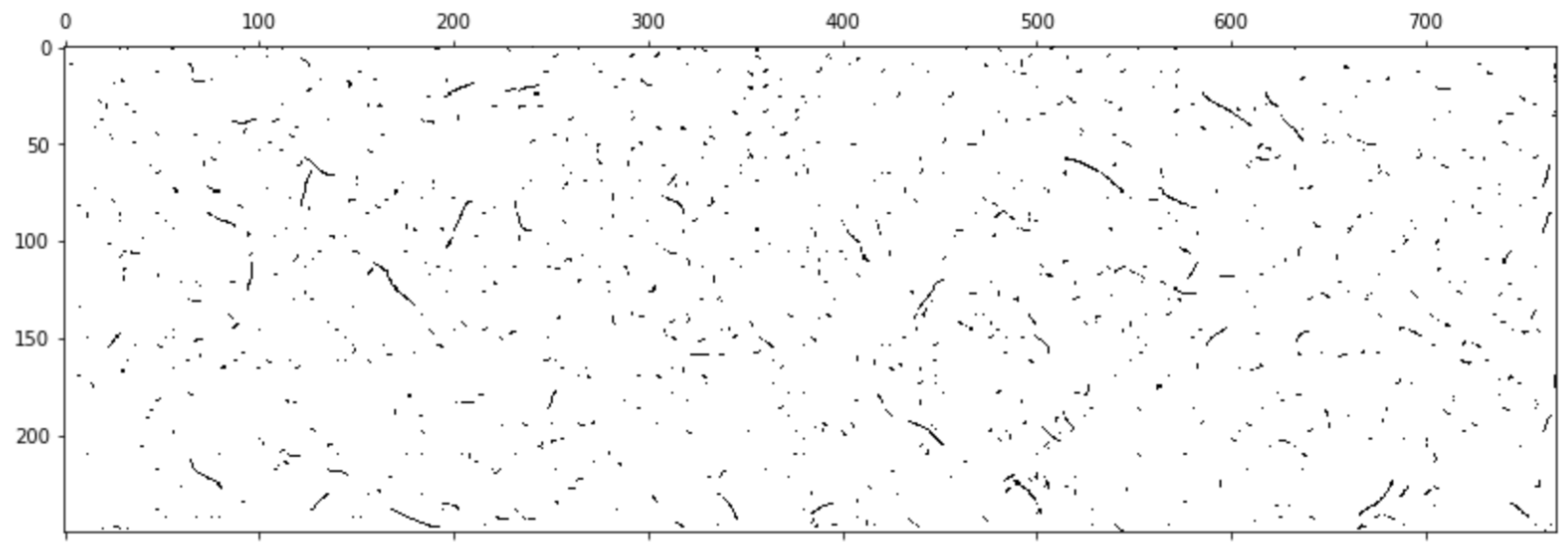} \label{fig:b}}
  \caption{Comparison of color-reversed PXD image (up) and generated image (down)} \label{fig-2}
\end{figure}

Fig.\ref{fig-2} shows a comparison between a generated hitmap and a real one for a sensor in the inner layer. One can generate sensor-dependent images for any layer and ladder number. It is very important for the model to pass the first qualitative evaluation test. Blurriness, image fidelity, general clustering details, and sharpness are some visual factors that one can check. Needless to mention that, we are also developing an automated and quantitative version of this evaluation, discussed in the Outlook section.

\begin{figure}[ht]
    \centering
    \subfloat[\centering Collapsed model]{{\includegraphics[width=9.5cm]{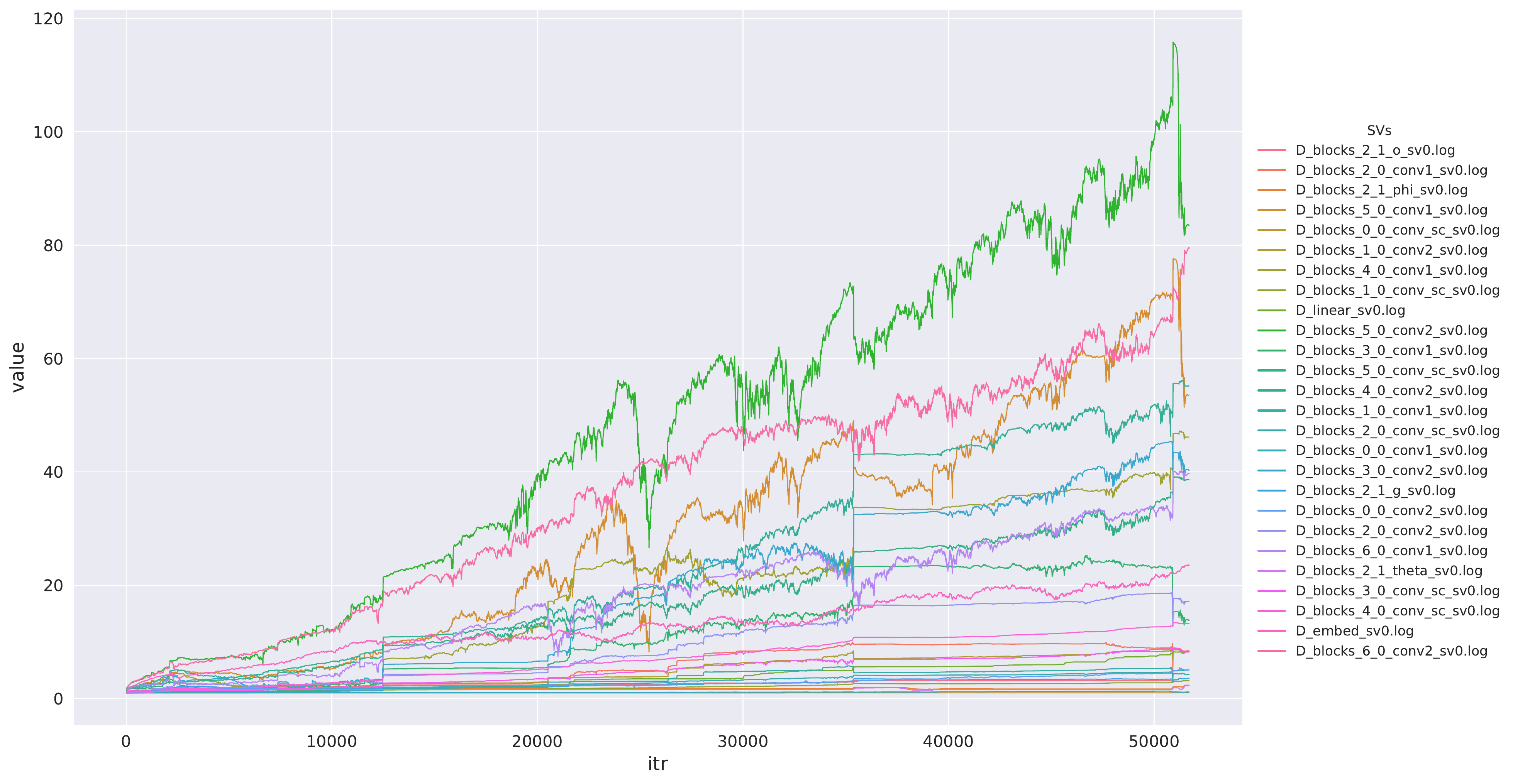} }}%
    \qquad
    \subfloat[\centering Our Model]{{\includegraphics[width=9.5cm]{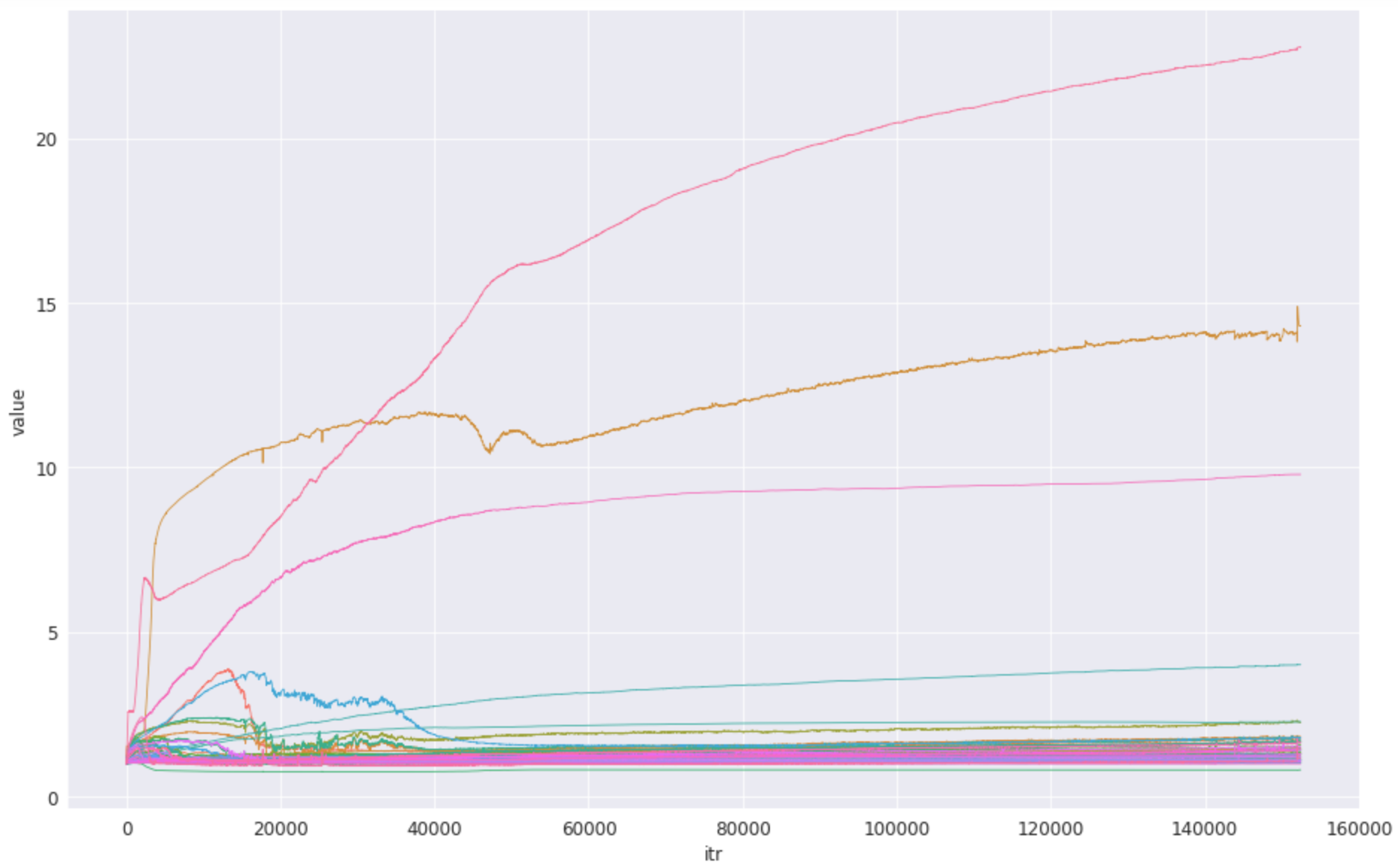}}}%
    \caption{First singular values of the discriminator for each layer in (a) a model with collapse (b) our model. The most important thing to notice here is the sudden drop in the values over a absolute large range which is an indication of discriminator's overfitting.}%
    \label{fig-3}%
\end{figure}

Regarding the first singular value of the weight matrix for different layers, our observation was in accord with \cite{biggan} that when the loss for the discriminator approaches zero constantly during training, it is due to the collapse. However, we also observed that for the discriminator, the spectra are very noisy, and when either mode collapse or discriminator memorization happens, the singular values drop suddenly. The most important thing to observe is the absolute change.
In Fig.\ref{fig-3}, one can see the first singular values of the weight matrices of each layer of the model. They can be efﬁciently computed using the Arnoldi iteration method~\cite{sv}. The first large drop and consequently collapse occurs around iteration 25000 with the green and orange layer which can be seen in Fig.\ref{fig-3}a.
As a result, the image contents gradually become complete nonsense. In order to address this issue and delay the collapse we tried and evaluated many methods such as Dropout, Orthogonal Regularization for both the generator and discriminator, Spectral regularization~\cite{specreg} and also gradient penalty~\cite{gp}. However, the gained performance was not desirable in terms of stability endurance. Then, after modifying the BigGAN-deep model with contrastive conditioning over the discriminator, we could achieve proper stability in the end.

\subsection{Validation Methods}
\label{sec-7}
\begin{figure}[ht]
\centering
\includegraphics[width=0.6\textwidth,clip]{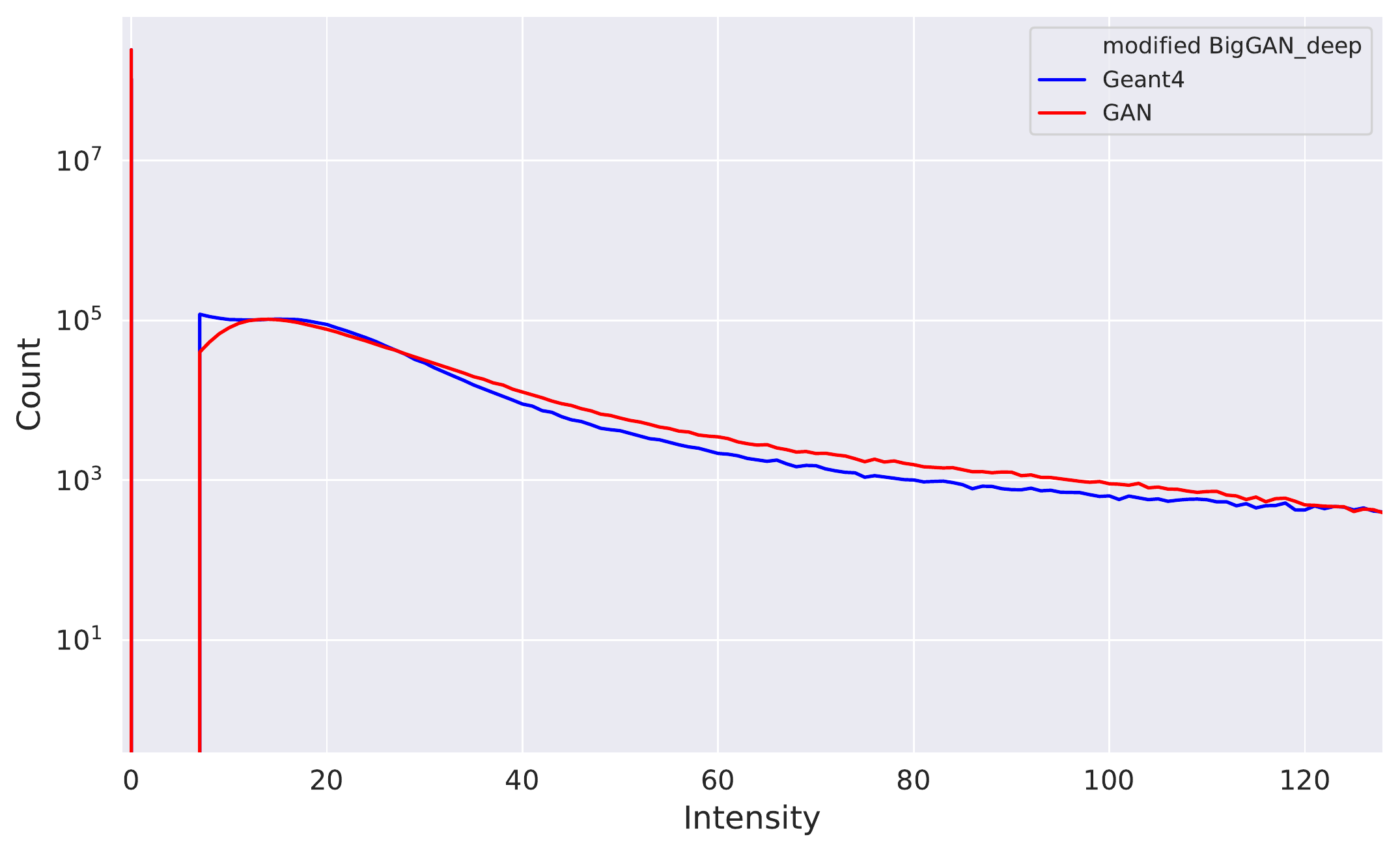}
\caption{Pixel intensity value comparison. Obviously a large portion of the images are totally dark. There is a very narrow discrepancy between the Geant4 simulated and GAN generated pixel values.}
\label{fig-4} 
\end{figure}

For the post-training validation process, we used 10000 new simulated events. First, we observed the pixel intensity values. Pixel value $0$ means complete blackness. Based on the MC simulation, there is a gap between 0 and 7 ADU as in fig.\ref{fig-4}. Below pixel intensity $7$, we used a function to cut off the residual pixel values. As a result, we will recover the separation between the two parts of the diagram. However, we will lose many pixel values which in principle correspond to a specific background process in the image. Moreover, the non-linearity of the intensity curve is not fully learned yet. We will address this issue in the discussion section.

\begin{figure}[ht]
\centering
\includegraphics[width=0.6\textwidth,clip]{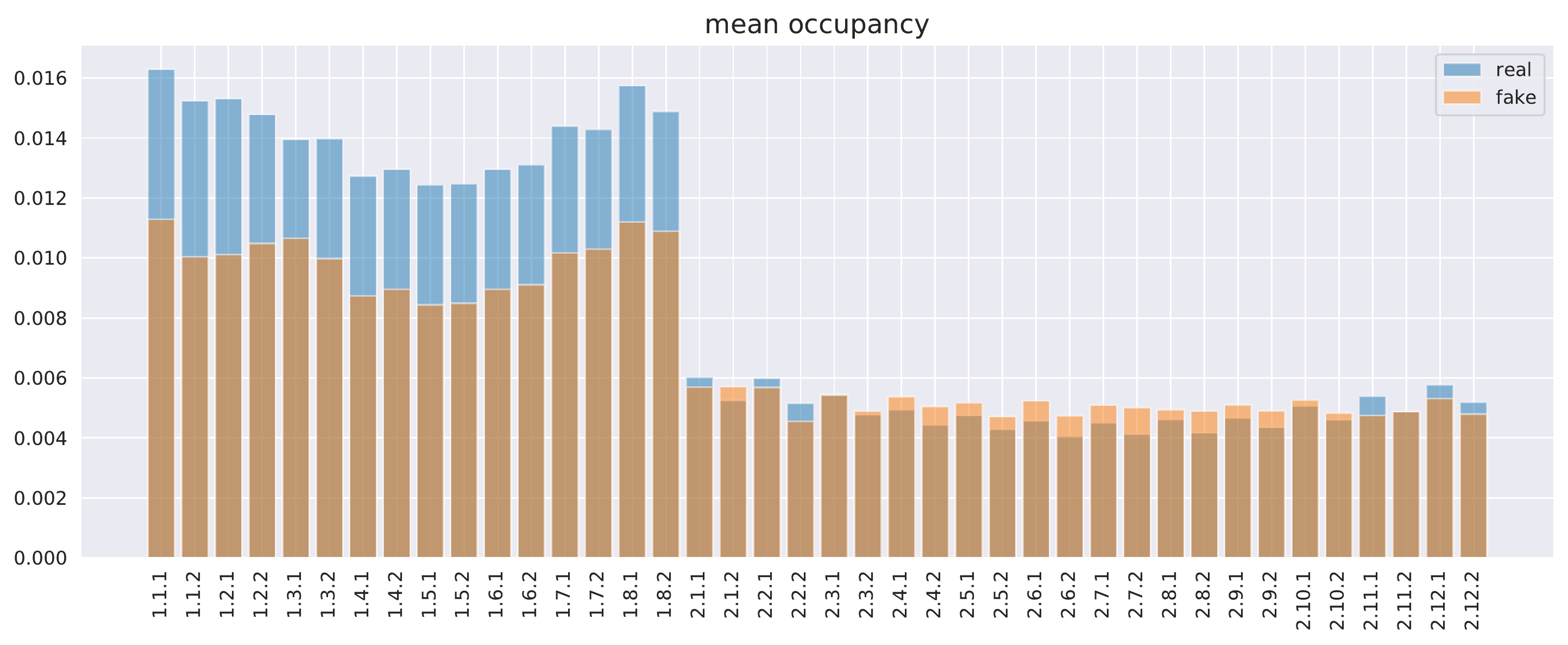}
\caption{Sensor dependent mean occupancy comparison between Geant4 simulated and GAN generated images}
\label{fig-5} 
\end{figure}

Our other metric for validation is the mean occupancy frequency comparison between the MC simulated test set and the generated images in Fig.\ref{fig-5}. This is a consequence of considering sensor IDs in our class-conditional GAN model. As a result of considering the labels, the model partially learned the frequency of PXD hits over each sensor which is of real importance. Subsequently, this indicates the generation of hitmaps closer to reality with better diversity and fidelity. The discrepancy between the real and generated mean occupancy over the hitmaps can be lowered by more tuning over the model. We propose more solutions in sec. \ref{sec-5}.

In the end, our last metric is a quantifiable comparison between the generated hitmaps and the simulated ones by studying the impact parameter resolutions. To do so, we simulated $\Upsilon(4S)$ signal events and considered three background overlay scenarios to compare with each other:
\begin{itemize}
    \item No background in the PXD
    \item Geant4 simulated PXD background
    \item Generated PXD background with modified BigGAN-deep
\end{itemize}

\begin{figure}[ht]
\centering
\includegraphics[width=0.6\textwidth, clip]{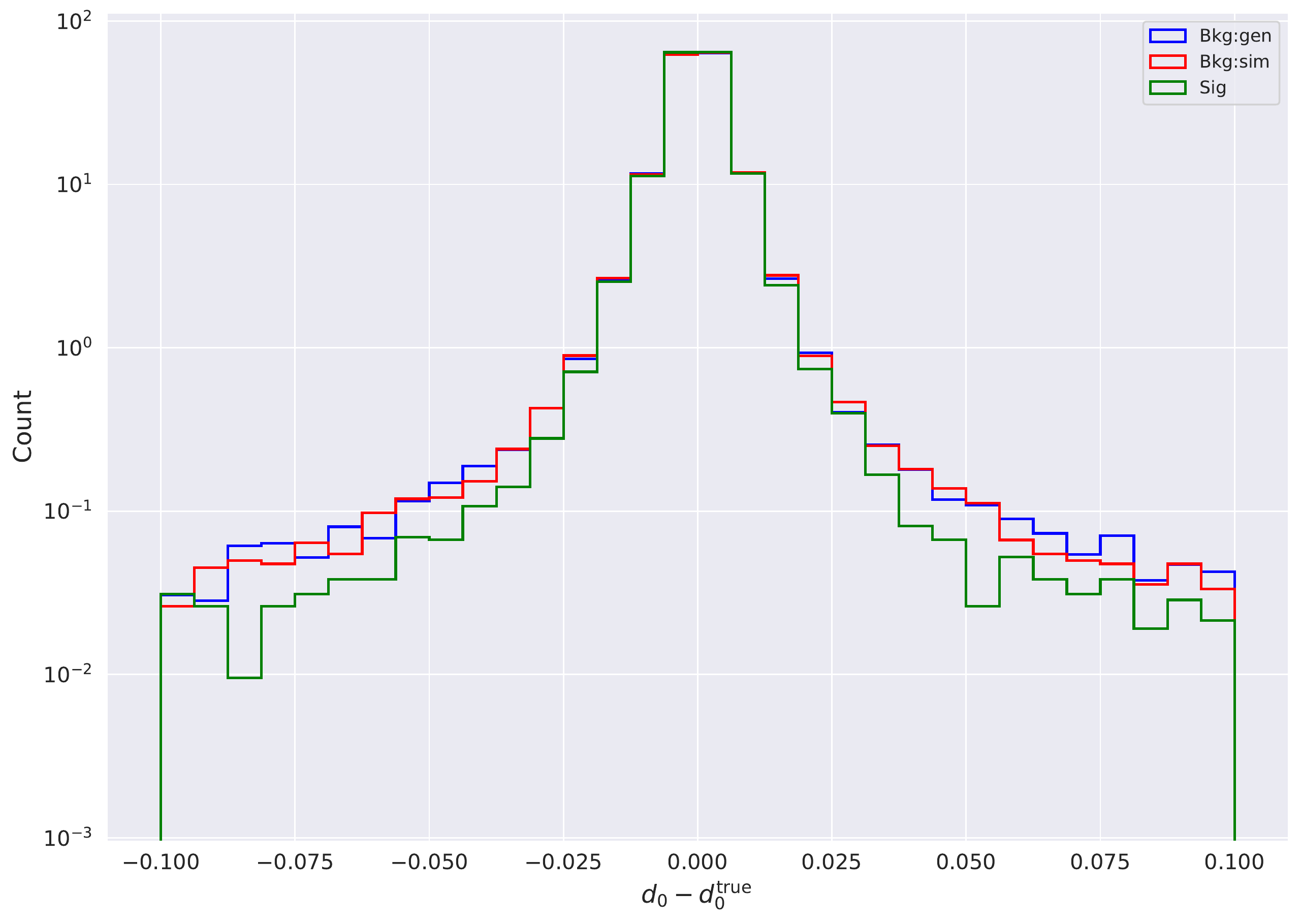}
\caption{Comparison of the resolution of the signed distance in the transverse plane from the pivotal point to the helix, between simulated and generated background and no background}
\label{fig-6} 
\end{figure}

The main effect of the background in the PXD is that wrong hits are attached to tracks. As a result, the effect of the background is visible over the tails of the distribution. The difference between the Geant4 simulated and conditional GAN is marginal and our model successfully reproduce Geant4 simulated background effects.
In order to quantify the difference between resolutions, we introduce the 2-point Wasserstein distance (Frechet Distance)~\cite{wass} which is a similarity measure for Gaussian distributions. For two multivariate Gaussian distributions with means $\mu_x$ and $\mu_y$ and covariance matrices, $\Sigma_x$ and $\Sigma_y$, the Frechet distance is defined as

\begin{equation}
    W^2(x,y)=|\mu_x-\mu_y|^2+\Tr(\Sigma_x+\Sigma_y-2\sqrt{(\Sigma_x\Sigma_y)}~).
\end{equation}

Looking at the two-point Wasserstein distance function between the distributions, the lower the Frechet Distance (FD), the better the image quality from the physics point of view. As an outcome of conditioning over the sensor IDs, the Frechet Distance score became even smaller in comparison to the former model.

\begin{table}[ht]
\centering
\begin{tabular}{ c|c c c } 
 \hline
  Model & Geant4. vs None. & GAN. vs None. & GAN. vs Geant4.  \\ 
 \hline
 \textbf{modified BigGAN-deep} & $5.46\mathrm{e}{-3}$ & $\mathbf{6.70\mathbf{e}{-3}}$ & $\mathbf{2.51\mathbf{e}{-3}}$ \\
 \hline
 WGAN-gp & $5.46\mathrm{e}{-3}$ & $6.98\mathrm{e}{-3}$ & $2.82\mathrm{e}{-3}$ \\
 \hline
 \hline
\end{tabular}
\caption{\label{tab:2}FD scores comparison between models in three background scenarios.}
\end{table}

\section{Summary and outlook}
\label{sec-5}
We developed a class-conditional GAN, based on PXD images and their sensor ID. As a result, we could achieve more reasonable agreement with real data and now are able to generate sensor-dependent PXD background hitmaps. Using GAN to generate the background data, we will reduce the storage consumption with the cost of additional CPU resources. In our path, we faced many challenges such as severe mode-collapse and memorization problem which was even more difficult in our case due to the lack of concrete training evaluation methods. We also did a vast ablation study over many state-of-the-art methods in the conditional image generation task and developed a stable modified GAN model.
\par
We are still trying to develop methods to overcome the remaining issues. For example, one idea is to add the prior information of pixel intensity values to the loss function of the generator like an L2 regularization scheme to penalize wrong intensity distribution in each batch. In other words, by minimizing the loss, the model will also minimize the L2 distance between the generator's normal output and the generator's output that has gone through pixel intensity cut-off. This will be in theory more effective than the hard cut-off method. We argue that if we only cut off this in a mere post-processing step, we will lose location-based data in the image for blackening only the pixel values. Since, despite the incorrectness of pixel values below 7 ADU, they correspond to would-be physical tracks which in principle we need to have them even with a lower intensity. 
\par
Another path worth trying would be to refine the GAN set up to capture the absolute relation between two layers of the PXD detector like a motion picture generation using sequence-based GAN models. Furthermore, as we direly need a reliable training metric, one can create a custom Inception Score (IS)~\cite{mbdis}, based on an image classification model trained on simulated events and images to predict PXD sensor IDs. The original Inception Score is based on the pre-trained Inception v3 model~\cite{insv3} with which a large number of generated fake images are classified. Consequently, the probability of the image belonging to each class is predicted. These predictions are then summarized into the Inception Score to have a fully automated evaluation metric for the training. Furthermore, the next step for an ultimate comprehensive validation of generated hitmaps would be an estimate of a systematic uncertainty on the tracking efficiency, fake rate, and resolution due using a GAN. If that is small compared to other systematic uncertainties we can use the model. Finally, as the PXD background generation is being integrated into the Belle II software~\cite{soft}, there will be many new experiences that may be very useful for other experiments as well.


\end{document}